\documentclass[preprint,showkeys,secnumarabic,showpacs,amsmath,amssymb]{revtex4}

\date{\today}

\begin{document}

\title{Generalized Brans-Dicke cosmology in the presence of matter and dark energy }

\author{Hossein Farajollahi}
\email{hosseinf@guilan.ac.ir} \affiliation{Department of Physics,
University of Guilan, Rasht, Iran}

\author{Narges Mohamadi}
\email{narges_mff@yahoo.com}

\affiliation{Department of Physics, University of Guilan, Rasht, Iran}

\begin{abstract}

We study the Generalized Brans-Dicke cosmology in the presence of matter and dark energy.
Of particular interest for a constant Brans-Dicke parameter, the de Sitter space has also been investigated.

\end{abstract}



\maketitle

\section{Introduction}

 The Brans-Dicke (BD) theory is defined by a scalar field $\varphi$ and
  a constant coupling function $\omega$ as perhaps
 the most natural extension of general theory of relativity
 that is obtained in the limit of $\omega\rightarrow\infty$ and
 $\varphi=$ constant. The theory appears
 naturally in supergravity theory, Kaluza-Klein theories and in all
 the known effective string actions. In FRW cosmology, the theory
 gives simple expanding solutions for scale factor $a(t)$ and scalar field
 $\varphi(t)$ which are compatible with solar
 system experiments \cite{Perlmutter}

The generalized BD
theory, sometimes referred to as graviton-dilaton or
scalar -tensor theory, is instead, defined by $\omega$ which is
 implicitly a function of time $\omega(t)$. Naturally, a few
attempts have been taken to study the dynamics of the universe using
this formalism \cite{Sahoo}\cite{Sahoo1}\cite{Sahoo2}.

The belief that modified gravity theories may have played a crucial
role during the early universe has recently been renewed by
extended inflation (for example see \cite{La}\cite{Guth}). In this case a scalar– tensor
gravity theory allows the first order phase transition of the "old"
inflationary model to complete. This arises because the
scalar field $\varphi$, that is essentially the inverse of the Newton's gravitational constant,
damps the rate of expansion and, in the original extended
inflationary model based on the BD theory, turns the
exponential expansion found in general relativity into power law
inflation \cite{Lucch}. However, BD theory is unable to
meet the simultaneous and distinct requirements placed by the
post–Newtonian solar system tests and by the need to
keep the sizes of the bubbles nucleated during inflation within the
limits permitted by the anisotropies of the microwave background
\cite{Reas} \cite{Steinhardt}.

In order to carry out a detailed study of the dynamics of the cosmic
evolution in this formalism, knowledge about exact time-dependence of
$a(t)$, $\varphi(t)$ and $\omega(t)$ and energy momentum distribution in spacetime is
essential. In part of this paper, similar to \cite{Diaz}, we have obtained red shift-dependence of
$\omega(z)$ with the power of red shift $z$ determined in terms of the
exponent of scale factor $a(t)$ which is taken to vary as
$a(t)\propto t^\delta$ and equation of state parameter for the matter
 and dark energy contribution. With the help
of observational evidence we obtain certain information about the
parameters describing the cosmological model in particular regarding
the early and late time behavior of the universe. We also investigate
 both empty and filled de Sitter space case with constant BD parameter
 and shows that the result is consistent with recent measurements.

\section{The Model}

 We consider a flat Universe filled with pressureless matter and dark
 energy both described by perfect fluid. The
 field equations in generalized BD
theory with time dependent $\omega$, are
\begin{equation}
  H^2\varphi^2+H\dot{\varphi}\varphi
  -\frac{\omega}{6}\dot{\varphi}^2=\frac{\rho_x+\rho_m}{3}\varphi,\label{fried-eq}
\end{equation}
\begin{equation}
   2\dot{H}\varphi^2+3H^2\varphi^2+\frac{\omega}{2}\dot{\varphi}^2+
   2H\dot{\varphi}\varphi+\ddot{\varphi}\varphi=-p_x\varphi,\label{acceler-eg}
\end{equation}
where $p_x=\alpha_x \rho_x$, $p_m=\alpha_m \rho_m$ are the equations
of state for dark energy and matter and the scale factor and scalar
field are respectively $a(t)$ and $\varphi(t)$. In addition, the
equation of motion for BD scalar field is given by
\begin{equation}
  \ddot{\varphi}+3H\dot{\varphi}=\frac{\rho_x+\rho_m-3p_x}{2\omega+3}
  -\frac{\dot{\omega}\dot{\varphi}}{2\omega+3}\label{scalar-eq}\cdot
\end{equation}
 From equations (\ref{fried-eq}), (\ref{acceler-eg}) and (\ref{scalar-eq}), the
energy conservation equation can be obtained as
\begin{equation}\label{4}
    (\dot{\rho}_x+\dot{\rho}_m)+3H(\rho_x+\rho_m+p_x)=0 \cdot
\end{equation}
Note that the wave equation for the BD scalar field,
(\ref{scalar-eq}), is not an independent expression as it follows
from the Bianchi identities alongside equations (\ref{fried-eq}) and
(\ref{acceler-eg}). In addition,  the dynamics of the scale factor
is governed not only by the matter and dark energy, but also by the
BD scalar field, $\varphi(t)$.

 One may assume that the matter and dark energy interact with each other,
  thus the growth of one is at the expense
of the other. Then the conservation equations for them are

\begin{eqnarray}
  &&\dot{\rho}_m+ 3H\rho_m= Q, \label{consx1}\\
  &&\dot{\rho}_x+ 3H(1+\alpha_x)\rho_x= -Q ,\label{consx}
\end{eqnarray}
where $Q>0$ stands for the interaction term. Alternately, one could construct
the equivalent uncoupled model described by:
\begin{eqnarray}
  \dot{\rho}_m+ 3H(1-\alpha_{m,eff})\rho_m&=& 0,\label{rhomeffect} \\
  \dot{\rho}_x+ 3H(1+\alpha_{x,eff})\rho_x&=&0,\label{rhoxeffect}
\end{eqnarray}
where
\begin{eqnarray}
  \alpha_{m,eff}=\frac{Q }{3H\rho_m}, \\
 \alpha_{x,eff}=\alpha_x+\frac{Q }{3H\rho_x}\cdot
\end{eqnarray}
The wave equation (\ref{scalar-eq}) is not altered by the
interaction equations (\ref{consx1}) and (\ref{consx}), since although the matter
and dark energy components do not conserve separately the overall
fluid -matter plus dark energy- does. Thus, one may introduce the
total energy density $\rho_{tot}=\rho_m+\rho_x$, and from equation
(\ref{rhomeffect}) and (\ref{rhoxeffect}), obtains
\begin{eqnarray}
  \dot{\rho}_{tot}+ 3H(1+\alpha_{tot})\rho_{tot}&=& 0, \label{tot-density}
\end{eqnarray}
with the solution
\begin{equation}
    \rho_{tot}\propto a^{-3(1+\alpha_{tot})},\label{density}
\end{equation}
where
\begin{eqnarray}
  \alpha_{tot}=\frac{p_x}{\rho_m+\rho_x}=\alpha_x\Omega_x \label{tot},
\end{eqnarray}
and $\Omega_x\equiv \frac{\rho_x}{\rho_{tot}}$. One can also find the rate of $\Omega_x$ as
\begin{eqnarray}
  \dot{\Omega}_{x}=\frac{-3H(\alpha_{x,eff}+\alpha_{m,eff})\rho_x \rho_m}{\rho^2_{tot}}
  =-\frac{3H\alpha_{tot}\rho_m+Q}{\rho_{tot}}, \label{tot-density1}
\end{eqnarray}
or in terms of red shift $z$,
\begin{eqnarray}
  \Omega '_{x}=\frac{3(\alpha_{x,eff}+\alpha_{m,eff})\rho_x \rho_m}{\rho^2_{tot}(1+z)}, \label{tot-densityz}
\end{eqnarray}
where $\frac{1}{a}=1+z$, and $a=1 $  is the present value of the
scale factor and $" ' "$ means derivative with respect to $z$.

Also in terms of the red shift $z$, equation (\ref{rhoxeffect}) can be rewritten
as
\begin{eqnarray}
  \rho '_x=\frac{3(1+\alpha_{x,eff})\rho_x}{1+z},\label{rhoxzeffect}
\end{eqnarray}
where the equation based on the sign of $1+\alpha_{x,eff}$
shows whether the density of dark energy will increase or not as
the red shift becomes low. For positive sign, the density decreases
like the quintessence, for negative sign, it increases like the
phantom, and when it is zero the density is invariant like the
cosmological constant.

\section{THE GENERAL $\omega$}

Following paper \cite{Sahoo} for $a \propto t^\delta$, $\varphi
\propto t^\beta$ and time dependent $\omega$, equation (\ref{fried-eq})
gives,
\begin{equation}
    \omega(t)= -\frac{2}{\beta^2}t^{-3\delta(1+\alpha_{tot})-\beta+2}\cdot \label{omega}
\end{equation}
One can rewrite equation (\ref{omega}) in terms of red shift $z$,
\begin{equation}
    \omega(z)= -\frac{2}{\beta^2}(1+z)^{3(1+\alpha_{tot})+(\beta-2)/\delta}, \label{omegaz}
\end{equation}
where its derivative with respect to $z$ is given by
\begin{equation}
    \omega'= \frac{2(-3\delta(1+\alpha_{tot})-\beta+2)}{\beta^2
    \delta}(1+z)^{2+3\alpha_{tot}+(\beta-2)/\delta}\cdot \label{omegaz-deriv}
\end{equation}
After some calculations, we can also rewrite equations (\ref{fried-eq}) and (\ref{scalar-eq}) as
\begin{equation}
  [(\frac{\dot{a}}{a}+\frac{\dot{\varphi}}{2\varphi})^2-\frac{(2\omega+3)\dot{\varphi}^2}{12\varphi^2}]3\varphi=\rho_{tot}, \label{fried-eq1}
\end{equation}
\begin{equation}
  \beta[\frac{3(1-\alpha_{tot})}{4}\beta+\frac{3\delta(1-\alpha_{tot})}{2}]=0.\label{scalar-eq1}
\end{equation}
From equation (\ref{scalar-eq1}) one
finds that for $\alpha_{tot}\neq 1$, $\beta$ restricted to be $0$ or  $-2\delta$. In case of $\alpha_{tot}=1$, there is no constraint on $\beta$. For $\beta=0$, from equation (\ref{omegaz}) one finds that $\omega\rightarrow\infty$ and from equation (\ref{fried-eq1}) we obtain $\varphi=\varphi_0=constant$ and $a\propto t^\delta$. So for $\beta =0 $, Brans- Dicke model goes over to General Relativity \cite{Alimi} and to obtain $\delta$, one has to solve
equations of General Relativity \cite{Peeb}. In case of
$\beta=-2\delta$, equations (\ref{omegaz}) and (\ref{omegaz-deriv})
reduce to,
\begin{equation}
   \omega(z)=-\frac{1}{2\delta^2}(1+z)^{(1+3\alpha_{tot})-2/\delta}\label{omegaz1},
\end{equation}
\begin{equation}
    \omega'= \frac{-\delta(1+3\alpha_{tot})+2}{2\delta^3}(1+z)^{3\alpha_{tot}-2/\delta}\cdot \label{omegaz-deriv1}
\end{equation}
It is clear from equations (\ref{omegaz1}) that, the parameter
$\alpha_{tot}$ which takes different values in different era,
controls the $z$ dependence of $\omega$ in different era. For today value of $z=0$, we have
\begin{equation}
    \omega_0=-\frac{1}{2\delta^2}\label{omegaz0},
\end{equation}
and
\begin{equation}
    \omega'_0= \frac{-\delta(1+3\alpha_{tot})+2}{2\delta^3}, \label{omegaz-deriv0}
\end{equation}
where for the present acceleration of the universe that $\delta$
needs to be greater than one, $\omega_0$ given by (\ref{omegaz0})
has the minimum value of $1/2$ in agreement with the observation
\cite{pavon}.

Further, in the last scattering surface, during the galaxy formation
era ($1<z<3$) where dark energy density must be sub-dominant to
matter density ($\alpha_{tot}>-0.5$), we have $-1/12<\omega<-1/14$.

In the Big Bang Nucleosynthesis (BBN) era where the presence of dark
energy should not disturb the observed Helium abundance in the
universe ($( \alpha_{tot})_{BBN} >- 0.21$ at $z = 10^{10}$)
\cite{Mathews}, we have $-9/32<\omega<-9/128$. This also shows that at
sometimes in the future, $z=-1$, we have
a big rip and $\omega \rightarrow -\infty$.

One also finds from (\ref{omegaz1}) and (\ref{omegaz-deriv1}) that
\begin{equation}
\frac{\omega'}{\omega}=((1+3\alpha_{tot})-2/\delta)(1+z)^{-1},
\end{equation}
where the ration for today is negative, in the distance future for $z=-1$
 it goes to minus infinity and in the distance past
where $z\rightarrow \infty$, it approaches zero.

\section{De sitter space time with constant $\omega$}

We now assume an empty de Sitter spacetime with the solution $H =
H_0$. Then, for $\varphi \propto e^\beta$ and constant BD parameter
$\omega$, equations (\ref{fried-eq}) and
(\ref{acceler-eg}) can be solved:
\begin{eqnarray}
 && H_0^2 +H_0 \beta-\frac{\omega}{6}\beta ^2=0, \\
 &&3H_0^2 +2H_0 \beta+(\frac{\omega}{2}+1)\beta ^2=0,
\end{eqnarray}
to give
\begin{eqnarray}
\beta&=&\frac{(-2\pm \sqrt{-8-6\omega})H_0}{2+\omega}, \\
 \beta&=&\frac{(3\pm \sqrt{9+6\omega})H_0}{\omega}\cdot
\end{eqnarray}
For these two solutions to be consistent implies that $\omega=-4/3$ or
$\omega=-3/2$. For equations  (\ref{fried-eq}), (\ref{acceler-eg})
and (\ref{scalar-eq}) to be simultaneously satisfied only $\omega=-4/3$ and so $\beta=-3H_0$ is acceptable. For
large value of $H_0$, during inflation, while the universe expands
exponentially, the BD scalar field drops exponentially.

In the presence of matter and dark energy we may also have a  de
Sitter solution $H = H_0$. Then the equations (\ref{fried-eq}) and
(\ref{acceler-eg}) become
\begin{eqnarray}
 && H_0^2 +H_0 \beta-\frac{\omega}{6}\beta ^2=\frac{1}{3}e^{[-3H_0(1+\alpha_{tot})-\beta]t}, \\
 &&3H_0^2 +2H_0 \beta+(\frac{\omega}{2}+1)\beta ^2=-\alpha_{tot}e^{[-3H_0(1+\alpha_{tot})-\beta]t}\cdot
\end{eqnarray}
These equations are satisfied when
\begin{eqnarray}
 && H_0^2 +H_0 \beta-\frac{\omega}{6}\beta ^2=\frac{1}{3}, \label{betaA}\\
 &&3H_0^2 +2H_0 \beta+(\frac{\omega}{2}+1)\beta ^2=-\alpha_{tot},\label{betaB}\\
 &&-3H_0(1+\alpha_{tot})=\beta  \label{betaC}\cdot
\end{eqnarray}
Using equation (\ref{betaC}) in (\ref{betaA}) one gets
\begin{eqnarray}
\omega
&=&\frac{2(-3H_0^2(2+3\alpha_{tot})-1)}{9H^2_0(1+\alpha_{tot})^2}\cdot
\end{eqnarray}
Similarly, equation (\ref{betaC}) in conjunction with equation
(\ref{betaB}) gives
\begin{eqnarray}
\omega &=&\frac{2(3H_0^2(1+2\alpha_{tot})-9H_0^2
(1+\alpha_{tot})^2-\alpha_{tot})}{9H_0^2(1+\alpha_{tot})^2}\cdot
\end{eqnarray}
For equations (\ref{betaA}), (\ref{betaB}) and (\ref{betaC}) to be
simultaneously satisfied, above two values of $\omega$ should be
equal. Imposition of this condition leads to
\begin{eqnarray}
\alpha_{tot} &=&\frac{-(3H_0^2+1)\pm
\sqrt{9H_0^4+42H_0^2+1}}{18H_0^2}\cdot
\end{eqnarray}
From the above solution we find that for negative sign and $H_0>10$,
  $\alpha_{tot}=-0.33$ or for $H_0<0.15$,  $\alpha_{tot}>-0.33$. For
  positive sign, for $H_0\gg1$ or $H_0\simeq 0$, we have  $\alpha_{tot}\simeq 0$.
  In case of $\alpha_{tot}>-0.33$ or from equation (\ref{tot}), $\Omega_x<
0.33$, this is consistent with last Scattering Surface, during the galaxy
formation era ($1 < z < 3$) where dark energy density must be sub-dominant
to matter density and accordingly $\Omega_x < 0.5$ . Then one gets
$\omega=-1.5$ and $\beta=-142.7$.

From the above argument as to the Brans-Dicke parameters $\omega$ are concerned,
it is negative and of the order of unity. This could be considered
as an unsatisfactory result, in view of the high lower limits imposed
to $\omega$ by astronomical tests in the Solar System.

A possible solution of this contradiction as discussed is in
considering a non-constant coupling function $\omega(t)$ in Generalised Brans-Dicke theory. Thus,
the value of such a function can change with the cosmic time and, in the limit
$t\rightarrow\infty$, it could agree with local measured values \cite{Will}. This argument is based on the scalar-tensor theories in which $\omega$ depends on the scale, being very high in the weak field approximation of
Solar System that probe only a limited range of space and time. To effectively
constrain more general scalar-tensor theories, one
would also like to have ''strong-field'' experiments, such as
that provided by the binary pulsar \cite{Damour}. It was also
pointed out that in cosmological models based on more
general scalar-tensor theories in which $\omega$ can vary, there is
generally an attractor mechanism that drives $\omega$ to $\infty$ at late
times \cite{Nordtvedt}.

\section{Conclusion}

In this work,we have derived the explicit time and red shift dependence of the Brans-Dicke parameter $\omega$
by solving gravitational field and wave equations of generalized BD
theory consistently in the present of matter and dark energy, assuming power law behavior for the scale
factor $a(t)$ and scalar field $\varphi(t)$. Similar to the work done in \cite{Sahoo}, we find
two consistent solutions of the field and wave equations. One solution leads to General Relativity and the other
one leads to a $z$-dependent
$\omega(z)$ whose red shift dependence is governed by the equation of state parameter
$\alpha_{tot}$. Consequently, $\omega(z)$ exhibits different
behavior in different epochs of the evolving Universe characterized
by its dominant matter/dark energy components. We also find that the ratio $\frac{\omega'}{\omega}$ is a negative monotonically increasing function of $z$. In
particular, for an expanding universe, we have studied the empty de Sitter space with constant $\omega$ and find that
in the era of inflation that $H_0$ is high, the scalar field drops exponentially and $\omega=-4/3$.
 Moreover, in the presence of matter and dark energy, we also find that $\omega$ is negative and we are able to explain the last scattering surface constraint

\end{document}